\documentclass[twocolumn,english,prd,tightenlines,floatfix,showpacs,preprintnumbers,nofootinbib,eqsecnum,superscriptaddress]{revtex4-1}
\pdfoutput=1

\usepackage{amsmath,color}
\usepackage{amssymb}
\usepackage{graphicx}
\usepackage{enumerate}
\usepackage{slashed}
\usepackage{hyperref}
\usepackage[utf8]{inputenc}

\newcommand{\dd}{\mathrm d}

\newcommand{\beq}{\begin{eqnarray}}
\newcommand{\eeq}{\end{eqnarray}}

\newcommand{\ptt}{\mathbf{\tilde{p}}}
\newcommand{\pt}{\mathbf{p}}
\newcommand{\kt}{\mathbf{k}}
\newcommand{\qt}{\mathbf{q}}

\def\independenT#1#2{\mathrel{\setbox0\hbox{$#1#2$}%
 \copy0\kern-\wd0\mkern4mu\box0}} 
\def\permil{\%\raise.10ex\hbox{$_{\scriptstyle 0}$}}
\def\bea{\begin{eqnarray}}
\def\eea{\end{eqnarray}}
\def\bsp{\begin{split}}
\def\esp{\end{split}}

\def\be{\begin{equation}}
\def\ee{\end{equation}}
\def\bea{\begin{eqnarray}}
\def\eea{\end{eqnarray}}
\def\bsp{\begin{split}}
\def\esp{\end{split}}

\begin{document}
%%%%%%%%%%%%%%%%%%%%%%%%%%%%%%%%%%%%%%%%%%%%%%%%%%%%%%
%%%%%%%%%%%%%%%%%%%%%%%%%%%%%%%%%%%%%%%%%%%%%%%%%%%%%%

%\preprint{IFJPAN-IV-2011-13}

\title{Transverse momentum dependent splitting functions at work: quark-to-gluon splitting} 

\author{M.\ Hentschinski}
% \email{}
\affiliation{Facultad de Ciencias F\'isico Matem\'aticas, Benem\'erita
             Universidad Aut\'onoma de Puebla, \\Puebla 1152, Mexico}
\affiliation{Instituto de Ciencias Nucleares, Universidad Nacional Aut\'onoma
             de M\'exico,\\ Apartado Postal 70-543, Cuidad de M\'exico 04510, Mexico}

\author{A. Kusina}
% \email{kusina@lpsc.in2p3.fr}
\affiliation{Laboratoire de Physique Subatomique et de Cosmologie\\
             53 Rue des Martyrs Grenoble, France}

\author{K. Kutak}
% \email{}
\affiliation{Instytut Fizyki J\c{a}drowej im. H. Niewodnicza\'nskiego,\\
             Radzikowskiego 152, 31-342 Krak\'ow, Poland}

\begin{abstract}
Using the recently obtained $P_{gq}$ splitting
function we extend the low $x$ evolution equation for
gluons to account for contributions originating from
quark-to-gluon splitting. In order to write down a
consistent equation we resum virtual corrections
coming from the gluon channel and demonstrate that
this implies a suitable regularization of the $P_{gq}$
singularity, corresponding to a soft emitted quark.
We also note that the obtained equation is in a
straightforward manner generalized to a nonlinear
evolution equation which takes into account effects
due to the presence of high gluon densities.
\end{abstract}

% \keywords{
% Splitting Functions, $k_T$-factorization, QCD evolution}

%%%%%%%%%%%%%%%%%%%%%%%%%%%%%%%%%%%%%%%%%%5

\maketitle
\tableofcontents
%%%%%%%%%%%%%%%%%%%%%%%%%%%%%%%%%%%%%%%%%%%%%%%%%%
%%%%%%%%%%%%%%%%%%%%%%%%%%%%%%%%%%%%%%%%%%%%%%%%%%

%%%%%%%%%%%%%%%%%%%%%%%%%%%%%%%%%%%%%%%
\section{Introduction}
%%%%%%%%%%%%%%%%%%%%%%%%%%%%%%%%%%%%%%%
Parton distribution functions (PDFs) provide essential input to
phenomenology at today's collider experiments. In combination with
partonic cross-sections, which can be systematically calculated within
QCD perturbation theory, they allow for a very accurate descriptions
of `hard' events in hadron-hadron and hadron-electron collisions,
where `hard' refers to the presence of a scale $M$ significantly
larger than typical hadronic scales of the order of
$\Lambda_{\text{QCD}} \sim 200$ MeV.  While the bulk of such analysis is carried
out within the framework of collinear factorization, there exist
classes of multi-scale processes where the use of more general schemes
is of advantage. Such schemes involve in general
Transverse-Momentum-Dependent (TMD)\footnote{For a review see
  \cite{Angeles-Martinez:2015sea}.} or `unintegrated' PDFs in contrast
to conventional PDFs defined within collinear factorization which
depend only on the hadron longitudinal momentum fraction carried by the parton.

A particularly interesting example of such a multi-scale process is
provided by the high-energy or low $x$ limit of hard processes
$s \gg M^2 \gg \Lambda_{\text{QCD}}^2$ where $\sqrt{s}$ denotes the
center-of-mass energy of the process and $x = M^2/s$.  In such a
scenario it is necessary to resum terms enhanced by logarithms
$\ln 1/x$ to all orders in the strong coupling constant $\alpha_s$,
which is achieved by the Balitsky-Fadin-Kuraev-Lipatov (BFKL)
\cite{Kuraev:1976ge,Balitsky:1978ic} evolution equation. The resulting
formalism called high energy factorization or $k_T$ factorization
\cite{Catani:1990eg,Catani:1990xk,Dominguez:2011wm,Kotko:2015ura}
provides then a factorization of such cross-sections into a TMD
coefficient \cite{Deak:2009xt,vanHameren:2012if,vanHameren:2012uj} or
`impact factor' and an `unintegrated' gluon density.\\

While well defined in the `low $x$' limit, the ensuing formalism and
evolution equation of the unintegrated gluon density is confronted
with difficulties if one attempts a na\"ive extension into the `large
or moderate $x$' region. In particular, this concerns implementations
of unintegrated parton densities in parton showers of Monte-Carlo
event generators, as well as  observables in
  hadron-hadron collisions and/or  combinations   with fragmentation functions
which involve integrals over the full $x$ range of initial state
PDFs. One of the most tantalizing deficits is the limitation to
gluon-to-gluon splittings in the low $x$ evolution with quarks being
absent. While well justified in the limit $x \ll 1$, this restriction
omits a resummation of collinear logarithms associated with quark
splittings which provide sizable contributions at intermediate and
large $x$.  Furthermore, some of the hard collision processes are
initiated by quarks \cite{Hautmann:2012rf} and therefore the
appropriate unintegrated parton density functions are needed
\cite{Hautmann:2014uua}. To overcome this limitation,
in~\cite{Gituliar:2015agu} the real parts of quark induced TMD
splitting functions have been calculated within $k_T$-factorization,
supplementing earlier results of Catani and
Hautmann~\cite{Catani:1994sq, Hautmann:2012sh} who calculated the TMD
gluon-to-quark splitting function. This calculation has been based on
an extension of the method of the classical work by Curci Furmanski
and Petronzio~\cite{Curci:1980uw}, formulated for the collinear
factorization.%
\footnote{The Curci Furmanski Petronzio scheme was recently modified
  in order to simplify the structure of infra-red singularities, and
  used for calculating inclusive as well as unintegrated NLO splitting
  functions for the purpose of MC simulations (these results are in
  the framework of collinear
  factorization)~\cite{Jadach:2011kc,Gituliar:2014eba,Jadach:2016zgk}.}
% and was later used in ref.~\cite{Catani:1994sq} to calculate
% the $P_{qg}$ splitting function within the high-energy factorization.
Together with the already long-known $k_T$-dependent $P_{gg}$
splitting
function~\cite{Ciafaloni:1987ur,Catani:1989sg,Catani:1989yc,Marchesini:1994wr}
it gives a complete set of (real) emission
evolution kernels within the $k_T$-factorization scheme.\\

One of the peculiarities of the quark-induced splitting kernels is
that they develop a singularity, which is associated with the
vanishing of the transverse momentum of the parton emitted during the
splitting. While in the case of the quark-to-quark splitting the
resulting singularity is expected to be cancelled by the corresponding
virtual corrections, the situation is less clear in the case of the
quark-to-gluon splitting, where such virtual correction is absent at
leading order.  While a complete treatment of this singularity
requires eventually the development of a suitable scheme which removes
this singularity, we will address in the following a different
question. Namely, whether it is possible to formulate evolution
equations which, due to their particular structure, regularize this
divergence.

To this end we formulate an evolution equation for the gluon
distribution which includes already the quark-to-gluon splitting and
therefore first corrections due to quarks. Besides of testing possible
implementation of the quark-to-gluon splitting function, this equation
will allow us -- even before the complete set of equations is known --
to investigate the impact of including quarks into the purely gluonic
picture given by BFKL like evolution equations.

We should also note that one of the possibilities to construct a set
of TMD parton distributions is to use the Kimber Martin Ryskin
framework~\cite{Kimber:1999xc,Kimber:2001sc}. In this approach one
starts from collinear PDFs and constructs the TMD PDFs by introducing
coherence effects in the last step of the evolution. The widely used
PDF sets constructed in this
manner
are used rather sucessfully in phenomenological studies~\cite{Schafer:2016qmk,Maciula:2015kea,Ducloue:2015jba,Nefedov:2013ywa,Baranov:2016mix}. However, the potential problem which one faces is the
limitation to rather large and moderate values of $x$ and also large
transversal momenta. Therefore, one can not address questions related
to gluon saturation and the impact of gluon saturation on quarks.\\

The rest of the paper is organized as follows. In
Sec.~\ref{sec:quark-splitt-funct} we remind the result for the
$P_{gq}$ splitting function obtained in~\cite{Gituliar:2015agu} and
adopt it to a form suitable for the current work. In
Sec.~\ref{sec:comb} we formulate a generalization of BFKL equation
including quark contributions from the $P_{gq}$ splitting and perform
partial resummation of this contribution. Next the numerical stability
of this new equation is studied. Finally we calculate its
  high energy limit and show that in this limit we obtain a full
  resummation of the quark part.  In Sec.~\ref{sec:conclusions} we
summarize the obtained results and give perspective for further
studies.

\section{The TMD quark-to-gluon splitting function }
\label{sec:quark-splitt-funct}

\begin{figure}[!t]
\centerline{
\includegraphics[height=5cm,width=3cm]{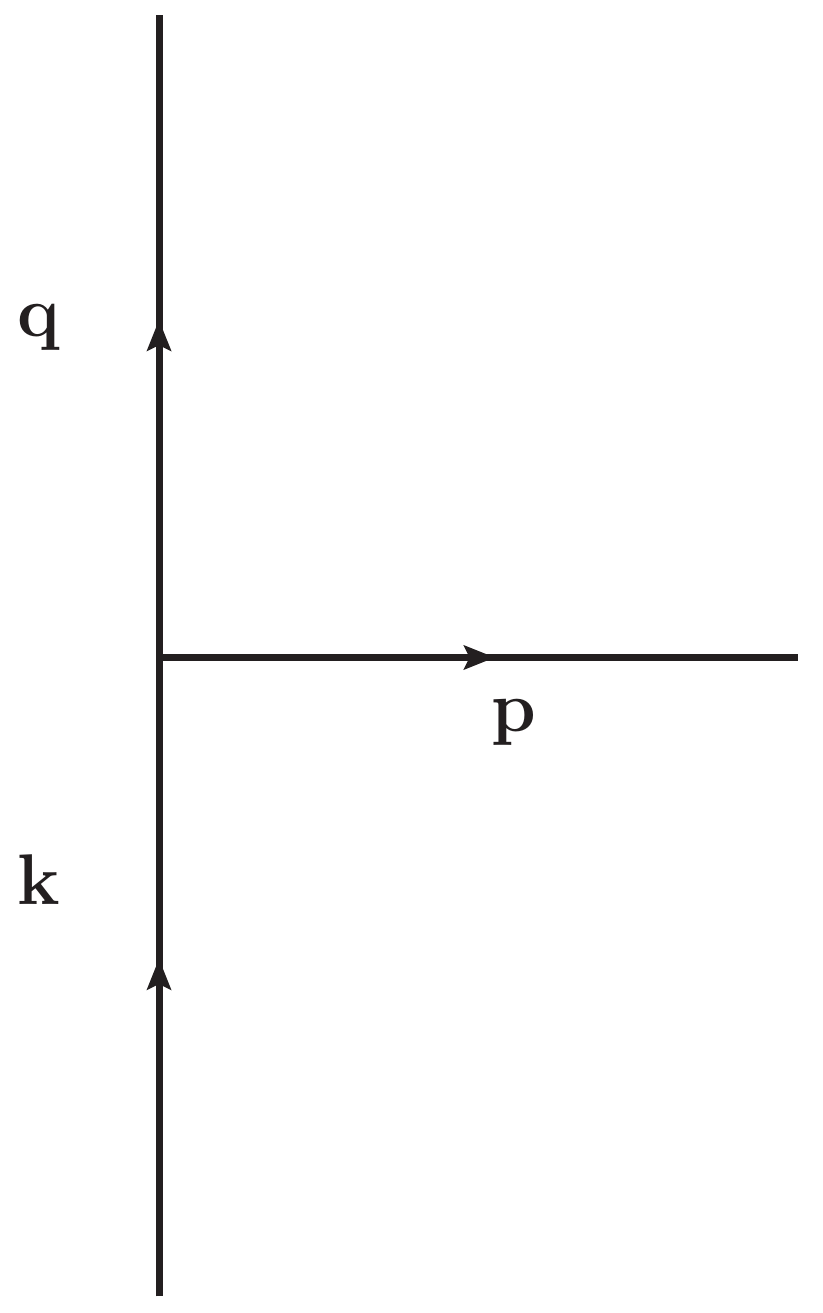}}
\caption{{\it An initial state parton with transverse momentum ${\bf k}$ splits into a parton with transverse momentum ${\bf q}$, while emitting a real parton with momentum ${\bf p} ={ \bf  k} - { \bf q}$.}}
\label{fig:Kinematics}
\end{figure}

We start by recalling the results of~\cite{Gituliar:2015agu}, where
the TMD quark-to-gluon splitting function $P_{gq}$ has been derived,
using an extension of the $k_T$-factorization formalism to initial,
off-shell quarks.  The corresponding kinematics is depicted in
Fig.~\ref{fig:Kinematics}.
%--------
%--------
The result of \cite{Gituliar:2015agu} is at first provided in terms of
the so-called TMD splitting kernel which allow to identify the
splitting functions of interest. Since we are in the following in particular interested
in isolating the singularity associated with the vanishing transverse
momentum $\pt$ of the real emitted quark we start with a
representation in terms of the rescaled transverse momentum
$\ptt=\frac{\kt-\qt}{1-z}$. The TMD splitting kernel is then given by
\begin{align}
  \label{eq:17gq}
&\hat K_{gq}\left(z, \frac{\kt^2}{\mu_F^2}, \alpha_s , \epsilon\right)=
\\&  \frac{\alpha_s}{2 \pi} z \int \frac{d^{2 + 2\epsilon} \ptt}{\pi^{1+\epsilon} \mu^{2 \epsilon}}\, e^{-\epsilon \gamma_E} \,  \Theta\left(\mu_F^2 - (1-z) (\ptt - \kt)^2 - z \kt^2 \right)
\notag \\
& \hspace{1.6cm}\Bigg\{
 \frac{(1-z)^{2\epsilon}}{z} \frac{2}{\ptt^2} 
+ \frac{\ptt^2 z(1-z)^{2 + 2\epsilon} (1 + \epsilon)}{ [(1-z) (\ptt - \kt)^2 + z \kt^2]^2} \notag \\
&  \hspace{4.2cm}
-
\frac{2 (1-z)^{1 + 2\epsilon}}{ (1-z) (\ptt - \kt)^2 + z \kt^2}
\Bigg\}.
\end{align}
For the purpose of the following sections,  we rewrite this result using the transverse  momenta of the emitted particle, $\pt$,
and the outgoing particle, $\qt$, while $\kt = \pt + \qt$. Furthermore, since we will formulate the evolution equation in $d=4$ dimensions, we set from now on  $\epsilon=0$. We find 
\begin{align}
  \label{eq:17gq}
&\hat K_{gq}\left(z, \frac{\qt^2}{\mu_F^2}, \alpha_s, 0 \right)= \nonumber
\\&  \frac{\alpha_s}{2 \pi} z \int \frac{d^{2 } \pt}{\pi^{} } \,  
\Theta\left(\mu_F^2 - \frac{z\,\pt^2 + (1-z)\,\qt^2}{1-z}\right)\frac{P_{gq}(z,\pt,{\bf q})}{\pt^2}
\end{align}
where
\begin{equation}
\label{eq:Pgq}
P_{gq}(z,\pt,{\bf q})=C_F \frac{\pt^4 z^2-2 \pt^2 \qt^2 (z-1) z+2 \qt^4 (1-z)^2}{z \left(\qt^2 (1-z)+\pt^2 z\right)^2}.
\end{equation}
In the limit $\pt \to 0$, the splitting function reduces to
\begin{align}
  \label{eq:1}
  \lim_{\pt \to 0} P_{gq}(z,\pt,{\bf q}) & = \frac{2 C_F}{z}
\end{align}
and as a consequence, Eq.~\eqref{eq:Pgq} appears to develop a
singularity in this limit. We stress that this singularity is not to
be confused with the conventional collinear singularity, which is
associated with the collinear gluon-to-quark splitting function. As
demonstrated in \cite{Gituliar:2015agu}, the collinear singularity is
associated with the collinear limit $|\pt|, |\qt| \gg |\kt|$.  In the
following section we will present a possible implementation of this
splitting function into an evolution equation for the unintegrated
gluon density, taking particular care to avoid the dangerous region
$|\pt| \to 0$.

%%%%%%%%%%%%%%%%%%%%%%%%%%%%%%%%%%%%%%%
\section{Combining quarks and gluons}
\label{sec:comb}
%%%%%%%%%%%%%%%%%%%%%%%%%%%%%%%%%%%%%%%
% {\it The $q$ has interpreatation ok $\bf k$ which is for instance used in formula 9.204 (page 263) in the book Barone and Predazzi and fig 9.19 (page 256) in the same book. Furthermore one has to think what to do with $\theta$ function which appears above.}

Our starting point is the leading order (LO) BFKL  equation  which describes evolution in $\ln 1/x$  for  the dipole amplitude in
the momentum space:
\begin{align}
\label{eq:bkversion1}
&{\cal F}(x,\qt^2)={\cal F}^{0}(x,\qt^2)
+\overline\alpha_s\int_x^1\frac{dz}{z}\int\frac{d^2\pt}{\pi \pt^2}
\\&\qquad
\times \left[{\cal F}(x/z,|{\bold q}+{\pt}|^2)-
\theta(\qt^2-\pt^2){\cal F}(x/z,\qt^2)\right]  \nonumber
\end{align}
where $\overline\alpha_s=\frac{C_{A}\alpha_s}{\pi}$.
This form is particularly useful  to promote the BFKL equation to a system of
equations for quarks and gluons. First of all let us consider  the equation for
gluons at LO which apart from   gluons receives also
contributions from quarks. Formally such an equation can be formulated as:
\begin{align}
{\cal F}(x,\qt^2)&={\cal F}^{0}(x,\qt^2) +\overline\alpha_s P_{BFKL}(z,\pt,\qt)\otimes{\cal F}(x/z,\qt^2) \nonumber
\\&
+\frac{\alpha_s}{2\pi}P_{gq}(z,\pt,\qt)\otimes{\cal Q}(x/z,\qt^2)
\end{align}
where ${\cal Q}(x/z,\qt^2)$ is the distribution of quarks and $P_{gq}$ is given by
eq.~\eqref{eq:Pgq}.
Introducing the resolution scale $\mu$ and decomposing the kernel of
the gluonic part of \eqref{eq:bkversion1} into a resolved real
emission part with $\pt^2>\mu^2$ and the unresolved part with
$\pt^2<\mu^2$, we obtain
\begin{align}
\label{eq:bkversion2}
{\cal F}(x,\qt^2)&={\cal F}^{0}(x,\qt^2)\\\nonumber
&+\overline\alpha_s\int_x^1\frac{dz}{z}\int\frac{d^2\pt}{\pi \pt^2}{\cal F}(x/z,|{\bold q}+{\pt}|^2)\theta(\pt^2-\mu^2)\\\nonumber
&+\overline\alpha_s\int_x^1\frac{dz}{z}\int\frac{d^2\pt}{\pi \pt^2}\big[{\cal F}(x/z,|{\bold q}+{\pt}|^2)\theta(\mu^2- \pt^2) \nonumber
\\&\qquad\qquad\qquad\qquad\quad -\theta(\qt^2-\pt^2){\cal F}(x/z,\qt^2)\big] \nonumber \\
&+\frac{\alpha_s}{2\pi}\int_x^1dz\int\frac{d^2\pt}{\pi \pt^2}P_{gq}(z,\pt,\qt){\cal Q}(x/z,|\pt+\qt|^2)\nonumber\,.
\end{align}
The integral over $\pt$ in the quark part is divergent and needs to be
regulated.  In the following we achieve this through introducing the same cut-off  $\mu$  as  used for  the gluonic part. Technically this is achieved through including  for the quark part a theta-function  $\theta(\pt^2-\mu^2)$.%
\footnote{Whereas, in the gluon case, this additional scale is just
  for technical convenience as $1/\pt^2$ is regularized by the virtual
  contribution, in the case of quarks, $\mu$ scale is really needed for
  regularizing the corresponding expression.}

A convenient way of formulating the resummation (or exponentiation) of
virtual and unresolved real emissions in the case of the gluon part
can be achieved by expressing the evolution equation by its Mellin
transform which is then further manipulated \cite{Kwiecinski:1996td}. The Mellin transform is
defined as
\begin{eqnarray}
\overline{\cal F}(\omega,\qt^2)&=\int_0^1 dx x^{\omega-1} {\cal F}(x,\qt^2), \nonumber
\\
\overline{\cal Q}(\omega,\qt^2)&=\int_0^1 dx x^{\omega-1} {\cal Q}(x,\qt^2)
\end{eqnarray}
while the inverse transform reads
\begin{eqnarray}
{\cal F}(x,\qt^2)&=\frac{1}{2\pi i} \int_{c-i\infty}^{c+i\infty} d\omega\, x^{-\omega} \overline{\cal F}(\omega,\qt^2),\nonumber
\\
{\cal Q}(x,\qt^2)&=\frac{1}{2\pi i} \int_{c-i\infty}^{c+i\infty} d\omega\, x^{-\omega} \overline{\cal Q}(\omega,\qt^2).
\label{eq:invmellin}
\end{eqnarray}
Performing the Mellin transform and using in the unresolved part $|{\bold q}+{\pt}|^2\approx{\bold q}^2$
(since $\pt^2<\mu^2$) we obtain
\begin{align}
\label{eq:eqtransform}
\overline {\cal F}(\omega,\qt^2)&=\overline {\cal F}^{0}(\omega,\qt^2)\\\nonumber
&+\frac{\overline\alpha_s}{\omega}\int\frac{d^2{\pt}}{\pi\pt^2}
[\overline {\cal F}(\omega,|\qt +\pt|^2)\theta(\pt^2-\mu^2)]\nonumber \\
&+\frac{\overline\alpha_s}{\omega}\int\frac{d^2{\pt}}{\pi\pt^2}{\overline {\cal F}}(\omega,\qt^2)
[\theta(\mu^2-\pt^2)-\theta(\qt^2-\pt^2)] \nonumber \\
&+\frac{\alpha_s}{2\pi}\int\frac{d^2{\pt}}{\pi\pt^2}\overline P_{gq}(\omega,\qt,\pt){\overline {\cal Q}}(\omega,|\qt+\pt|^2)
\theta(\pt^2-\mu^2) \nonumber
\end{align}
where
\begin{equation}
\overline P_{gq}(\omega,\pt,\qt)=\int_0^1 dz P_{gq}(z,\pt,\qt) z^{\omega}.
\end{equation}
For the moment the Mellin transform of the quark part is treated as
a formal expressions giving a convenient short-hand notation.  After
combining the unresolved real and virtual parts of the gluonic terms
we obtain
%----------------
\begin{widetext}
\begin{align}
\overline {\cal F}(\omega,\qt^2)&=\overline {\cal F}^{0}(\omega,\qt^2)\\\nonumber
&+\frac{\overline\alpha_s}{\omega}\int\frac{d^2\pt}{\pi \pt^2}\overline {\cal F}(\omega,|\qt +\pt|^2)\theta(\pt^2-\mu^2)
-\frac{\overline\alpha_s}{\omega}\overline {\cal F}(\omega,\qt^2)\ln\frac{\qt^2}{\mu^2}\\\nonumber
&+\frac{\alpha_s}{2\pi}\int\frac{d^2{\pt}}{\pi\pt^2}P_{gq}(\omega,\pt,\qt)
{\overline {\cal Q}}(\omega,|\qt +\pt|^2)\, \theta(\pt^2-\mu^2).
\nonumber
\end{align}
With
\begin{equation}
\hat {\cal F}^{0}(\omega,\qt^2)=\frac{\omega}{\omega+\overline\omega}\,\overline{\cal F}^{0}(\omega,\qt^2),
\qquad
\overline\omega=\overline\alpha_s\ln\frac{\qt^2}{\mu^2}.
\end{equation}
This can be simplified to 
\begin{align}
\label{eq:resummed_Mellin}
\overline {\cal F}(\omega,\qt^2)&=\hat {\cal F}^{0}(\omega,\qt^2)
 \\&
+\frac{\overline\alpha_s}{\overline\omega+\omega}\int\frac{d^2\pt}{\pi \pt^2}
\overline {\cal F}(\omega,|\qt +\pt|^2)\, \theta(\pt^2-\mu^2)
\notag  \\&
+\frac{\alpha_s}{2\pi}\frac{\omega}{\omega+\overline\omega}\int\frac{d^2{\pt}}{\pi\pt^2}
P_{gq}(\omega,\pt,\qt){\overline {\cal Q}}(\omega,|\qt
     +\pt|^2)\, \theta(\pt^2-\mu^2) \notag
\end{align}
which is easily verified through multiplying
eq.~\eqref{eq:resummed_Mellin} by a factor
$\frac{\bar{\omega}+\omega}{\omega}$.

Now we will transform this expression back to $x$ space. We will do this
separately for the gluon and quark parts. We start by writing the
formal expression
\begin{align}
\label{eq:Mellin_g_ini}
{\cal F}(x,\qt^2)
&=
\frac{1}{2\pi i}\int_{c-i\infty}^{c+i\infty} d\omega\,x^{-\omega}\hat{\cal F}^{0}(\omega,\qt^2)
\\ &
\label{eq:Mellin_g}
+\frac{1}{2\pi i}\int_{c-i\infty}^{c+i\infty} d\omega\,
x^{-\omega} \frac{\overline\alpha_s}{\omega+\overline\omega}
\int\frac{d^2\pt}{\pi \pt^2}\int_0^1 dy\,y^{\omega-1}{\cal F}(y,|\qt+\pt|^2)] \theta(\pt^2-\mu^2)
 \\ &
\label{eq:Mellin_q}
+\frac{1}{2\pi i}\int_{c-i\infty}^{c+i\infty} d\omega\,
x^{-\omega}\frac{\alpha_s}{2\pi}\frac{\omega}{\omega+\overline\omega}
\int\frac{d^2{\pt}}{\pi\pt^2}\,
% \overline P_{gq}(\omega,\pt,\qt)
\int_0^1 dz z^{\omega} P_{gq}(z,\pt,\qt)
 \\   &\qquad\qquad\qquad\qquad\qquad\qquad\times
\int_0^1\,dyy^{\omega-1}{\cal Q}(y,|\qt+\pt|^2)\theta(\pt^2-\mu^2). \notag
\end{align}

%===============================
\subsection{Gluon part}
%===============================
For the part with the initial gluon distribution, eq.~\eqref{eq:Mellin_g_ini},
we merely define
\begin{equation}
\tilde{\cal F}^0(x,\qt^2)\equiv \frac{1}{2\pi i}\int_{c-i\infty}^{c+i\infty}
d\omega\,x^{-\omega}\hat{\cal F}^{0}(\omega,\qt^2).
\end{equation}
To transform the remaining gluonic part given by eq.~\eqref{eq:Mellin_g} it is
convenient to first note that
\begin{align}
\label{eq:ReggeFactor}
\frac{1}{\omega + \overline{\omega}} & =   \int_0^1 \frac{dz}{z} z^{\omega+ \overline{\omega}} \, .
\end{align}
Using this identity we find that the gluonic term transforms into
\begin{align}
& \frac{1}{2\pi i}\int_{c-i\infty}^{c+i\infty} d\omega\,
x^{-\omega} \,
\overline\alpha_s \int_0^1 \frac{dz}{z} z^{\omega+ \overline{\omega}} 
\int\frac{d^2\pt}{\pi \pt^2}\int_0^1 dy\,y^{\omega-1}{\cal F}(y,|\qt+\pt|^2)]\, \theta(\pt^2-\mu^2)
\\& = 
\overline{\alpha}_s 
\int \frac{d^2 \pt}{\pi \pt^2}
  \int_0^1 \frac{dz}{z} z^{ \overline{\omega}}  
  \int_0^1 \frac{d y}{y}   {\cal F}(y,|\qt+\pt|^2) \theta(\pt^2-\mu^2) \delta\left(\frac{zy}{x}-1 \right)
\notag \\
& =\overline{\alpha}_s 
\int \frac{d^2 \pt}{\pi \pt^2}
  \int_x^1 \frac{dz}{z}  \Delta_R(z, \qt^2, \mu^2)
    {\cal F}\left(\frac{x}{z},|\qt+\pt|^2\right)\theta(\pt^2-\mu^2),
\label{eq:gluonPart}
\end{align}
where we have introduced  the  Regge formfactor
\begin{equation}
\Delta_R(z,\qt^2,\mu^2)\equiv\exp\left(-\overline\alpha_s\ln\frac{1}{z}\ln\frac{\qt^2}{\mu^2}\right)
\end{equation}
which ensures that the gluonic part is well-behaving when $\mu^2\to0$
and therefore allows us to take $\pt^2\to0$.

%===============================
\subsection{Quark part}
%===============================
Now we turn to the quark part given by formula~\eqref{eq:Mellin_q}.
We note that for a generic function $f(\omega)$ we have 
\begin{align}
\label{eq:omega_der}
\int \frac{d\omega}{ 2 \pi i}  \omega  x^{-\omega} f(\omega) & = -
x \partial_x  \int \frac{d\omega}{ 2 \pi i}    x^{-\omega} f(\omega) .
\end{align}
We additionally isolate the  part singular in $z$ of the $P_{gq}$ kernel
\begin{align}
\label{eq:reduced_splitting}
P_{gq}(z, \pt, \qt) & \equiv \frac{1}{z} \cdot   \tilde{P}_{gq}(z, \pt, \qt) \, . 
\end{align}
This allows us to rewrite eq.~\eqref{eq:Mellin_q} as
\begin{align}
\label{eq:quark}
& -x \partial_x 
\frac{\alpha_s}{2 \pi} \int \frac{d\omega}{ 2 \pi i}  x^{-\omega}
  \int_0^1 \frac{dz_1}{z_1} z_1^{\omega+ \overline{\omega}}   
\int \frac{d^2 \pt}{\pi \pt^2}
\int_0^1 \frac{d z_2}{z_2}  z_2^\omega  \tilde{P}_{gq}(z_2, \pt, \qt)
\notag \\
& \hspace{7cm} \cdot 
  \int_0^1 d y y^{\omega-1} {\cal Q}(y,|\qt+\pt|^2)\theta(\pt^2-\mu^2)
\notag \\&
= -x \partial_x 
\frac{\alpha_s}{2 \pi}
\int \frac{d^2 \pt}{\pi \pt^2}
  \int_0^1 \frac{dz_1}{z_1} z_1^{ \overline{\omega}} 
\int_0^1 \frac{d z_2}{z_2}   \tilde{P}_{gq}(z_2, \pt, \qt) 
\notag \\
& \hspace{5cm} \cdot 
  \int_0^1 \frac{d y}{y}   {\cal Q}(y,|\qt+\pt|^2)\theta(\pt^2-\mu^2) \delta\left(\frac{z_1 z_2 y}{x}-1 \right)
\notag \\ &
= -x \partial_x 
\frac{\alpha_s}{2 \pi}
\int \frac{d^2 \pt}{\pi \pt^2}
  \int_0^1 \frac{dz_1}{z_1} z_1^{ \overline{\omega}}  \int_0^1 \frac{d y}{y}
  \theta(y z_1 - x) \tilde{P}_{gq}\left(\frac{x}{y z_1}, \pt, \qt \right) 
    {\cal Q}(y,|\qt+\pt|^2)\theta(\pt^2-\mu^2).
\end{align}
Next note that
\begin{align}
\label{eq:der}
-x \partial_x   \theta(y z_1 - x) & \tilde{P}_{gq}\left(\frac{x}{y z_1}, \pt, \qt \right)
\notag \\ 
&= 
x\delta(x - y z_1)  \tilde{P}_{gq}\left(1, \pt, \qt \right)
- \theta(y z_1 - x) \left[ \tilde{P}'_{gq}\left(\frac{x}{y z_1}, \pt, \qt \right) \frac{x}{y z_1}  \right],
\end{align}
where
\begin{align}
\tilde{P}'_{gq}\left(z, \pt, \qt \right) \equiv \frac{d}{dz}  \tilde{P}_{gq}\left(z, \pt, \qt \right)
= -C_F\frac{2(1-z)\pt^2\qt^4}{\left(z\pt^2+(1-z)\qt^2\right)^3}.
\end{align}
Now using eq.~\eqref{eq:der} we obtain for the quark contribution 
\begin{align}
  \label{eq:quark_cont}
%   \text{ Eq.~\eqref{eq:quark}} & = 
&
\frac{\alpha_s}{2 \pi}
\int \frac{d^2 \pt}{\pi \pt^2}
 \int_x^1 \frac{d z}{z}  z^{ \overline{\omega}} 
 \tilde{P}_{gq}\left(1, \pt, \qt\right) 
    {\cal Q}\left(\frac{x}{z},|\qt+\pt|^2\right)\theta(\pt^2-\mu^2)
\notag \\
&
-\frac{\alpha_s}{2 \pi}
\int \frac{d^2 \pt}{\pi \pt^2}
 \int_x^1 \frac{d z}{z}   \int_z^1 \frac{dz_1}{z_1} z_1^{ \overline{\omega}} 
 \left[ \tilde{P}'_{gq}\left(\frac{z}{z_1}, \pt, \qt\right)  \frac{z}{z_1} \right]
    {\cal Q}\left(\frac{x}{z},|\qt+\pt|^2\right)\theta(\pt^2-\mu^2).
\end{align}
Note that $ \tilde{P}_{gq}\left(1, \pt, \qt\right)  = C_F$.
While the convolution integral over $z$ involves the
(non-perturbative) unintegrated quark density, the $z_1$ integral (in the second line) can
in principle be calculated analytically.

%===============================
\subsection{Combined gluon and quark parts}
\label{subsec:comb}
%===============================
Combining the results for the gluon part~\eqref{eq:gluonPart} and quark
part~\eqref{eq:quark_cont} we obtain
\begin{align}
{\cal F}(x,\qt^2)
&=
\tilde{\cal F}^0(x,\qt^2)
\nonumber\\&+
\overline{\alpha}_s 
\int \frac{d^2 \pt}{\pi \pt^2}
  \int_x^1 \frac{dz}{z}  \Delta_R(z, \qt^2, \mu^2)
    {\cal F}\left(\frac{x}{z},|\qt+\pt|^2\right)\theta(\pt^2-\mu^2)
\nonumber\\&+
\frac{\alpha_s}{2 \pi}
\int \frac{d^2 \pt}{\pi \pt^2}
 \int_x^1 \frac{d z}{z}  z^{ \overline{\omega}} 
 \tilde{P}_{gq}\left(1, \pt, \qt\right) 
    {\cal Q}\left(\frac{x}{z},|\qt+\pt|^2\right)\theta(\pt^2-\mu^2)
\nonumber\\
&
-\frac{\alpha_s}{2 \pi}
\int \frac{d^2 \pt}{\pi \pt^2}
 \int_x^1 \frac{d z}{z}   \int_z^1 \frac{dz_1}{z_1} z_1^{ \overline{\omega}} 
 \left[ \tilde{P}'_{gq}\left(\frac{z}{z_1}, \pt, \qt\right)  \frac{z}{z_1} \right]
    {\cal Q}\left(\frac{x}{z},|\qt+\pt|^2\right)\theta(\pt^2-\mu^2).
\end{align}
Note that 
$z^{ \overline{\omega}} = \Delta_R(z, \qt^2, \mu^2)$. Using in
addition 
$ \tilde{P}_{gq}\left(1, \pt, \qt\right)  = C_F$, we obtain 
\begin{align}
\label{eq:linear}
 {\cal F}(x,\qt^2)
& =
\tilde{\cal F}^0(x,\qt^2)
\notag \\ &    +
\frac{\alpha_s}{2 \pi}
\int_x^1 \frac{dz}{z}  \int \frac{d^2 \pt}{\pi \pt^2} \theta(\pt^2-\mu^2)   \Bigg[  
\notag \\
& \hspace{3cm}
    \Delta_R(z, \qt^2, \mu^2)
 \bigg( 2 C_A
    {\cal F}\left(\frac{x}{z},|\qt+\pt|^2\right)
+ C_F 
    {\cal Q}\left(\frac{x}{z},|\qt+\pt|^2\right) \bigg)
\nonumber\\
& \hspace{2.2cm}
-
     \int_z^1 \frac{dz_1}{z_1}  \Delta_R(z_1, \qt^2, \mu^2)
 \left[ \tilde{P}'_{gq}\left(\frac{z}{z_1}, \pt, \qt\right)
             \frac{z}{z_1} \right] 
  {\cal Q}\left(\frac{x}{z},|\qt+\pt|^2\right) \Bigg]
\end{align}
From the above expression we can see that the $\frac{1}{\pt^2}$
singularity of the quark term in the third line of
eq.~\eqref{eq:linear} is regularized by the Regge formfactor in direct
analogy with the gluonic term. For the term in the fourth line we note
on the other hand that $\tilde{P}_{gq}'(z, \pt, \qt) \sim \pt^2$ for
finite $\qt$; the limit $\mu^2\to 0$ is therefore expected to be
finite for this term.  Before addressing the numerical stability in
the next section, we note that the above resummation can be in a
straight forward manner extended \cite{Kutak:2011fu,Kutak:2012qk} to
the situation where the gluon density is large and therefore subject
to a nonlinear evolution equation, taking into account saturation
effects \cite{Gribov:1984tu,Gelis:2010nm}.  The extended non-linear
equation reads:
\begin{align}
\label{eq:nonlin}
 {\cal F}(x,\qt^2)
& =
\tilde{\cal F}^0(x,\qt^2) +
\frac{\alpha_s}{2 \pi}
\int_x^1 \frac{dz}{z}  \int \frac{d^2 \pt}{\pi \pt^2} \theta(\pt^2-\mu^2)   \Bigg[
\notag \\ &      
    \Delta_R(z, \qt^2, \mu^2)
 \Bigg\{ 2 C_A
    {\cal F}\left(\frac{x}{z},|\qt+\pt|^2\right)
+ C_F 
    {\cal Q}\left(\frac{x}{z},|\qt+\pt|^2\right)
\notag \\
&
\hspace{-1.1cm}-  \frac{ 4\pi \alpha_s}
{ R^2}\pt^2\delta(\pt^2-\qt^2)\left[
\left(
\int_{{\pt}^2}^{\infty} \frac{\dd l^2}{l^2}
{\cal F}\left(\frac{x}{z}, l^2\right)
\right)^2
+
{\cal F}\left(\frac{x}{z}, \pt^2\right)\int_{\pt^2}^{\infty} \frac{\dd l^2}{l^2}
\ln \left(\frac {l^2}{{\pt}^2}\right)
{\cal F}\left(\frac{x}{z}, l^2\right)
\right]
 \Bigg\}
\nonumber\\
& \hspace{2.5cm}
-
     \int_z^1 \frac{dz_1}{z_1}  \Delta_R(z_1, \qt^2, \mu^2)
 \left[ \tilde{P}'_{gq}\left(\frac{z}{z_1}, \pt, \qt\right)
             \frac{z}{z_1} \right] 
  {\cal Q}\left(\frac{x}{z},|\qt+\pt|^2\right) \Bigg]
% {\cal F}(x,\qt^2)
% &=
% \tilde{\cal F}^0(x,\qt^2)
% \\&+
%
% \overline{\alpha}_s 
% \int \frac{d^2 \pt}{\pi \pt^2}
%   \int_x^1 \frac{dz}{z}  \Delta_R(z, \qt^2, \mu^2)
%     {\cal F}\left(\frac{x}{z},|\qt+\pt|^2\right)\theta(\pt^2-\mu^2)
% \nonumber\\&+
% \frac{\alpha_s}{2\pi}C_F
% \int \frac{d^2 \pt}{\pi \pt^2}
%  \int_x^1 \frac{d z}{z}  \Delta_R(z, \qt^2, \mu^2)
%     {\cal Q}\left(\frac{x}{z},|\qt+\pt|^2\right)\theta(\pt^2-\mu^2)
% \nonumber\\&
% -\frac{\alpha_s}{2 \pi}
% \int \frac{d^2 \pt}{\pi \pt^2}
%  \int_x^1 \frac{d z}{z}   \int_z^1 \frac{dz_1}{z_1}  \Delta_R(z_1, \qt^2, \mu^2)
%  \left[ \tilde{P}'_{gq}\left(\frac{z}{z_1}, \pt, \qt\right)  \frac{z}{z_1} \right]
%     {\cal Q}\left(\frac{x}{z},|\qt+\pt|^2\right)\theta(\pt^2-\mu^2).
%\nonumber\\&
 % \int_x^1\frac{dz}{z}\Delta_R(z, \qt^2, \mu^2)
           %
\end{align}
where the gluonic part is given by momentum space formulation
\cite{Kutak:2003bd,Bartels:2007dm} of the Balitsky-Kovchegov equation
\cite{Kovchegov:1999yj,Balitsky:1995ub}.  In the above equation the
parameter $R$ has an interpretation of radius of the hadron.
\end{widetext}
%----------------

%===============================
\subsection{Numerical studies} 

\begin{figure*}[th]
\centering
  \includegraphics[width=0.46\textwidth]{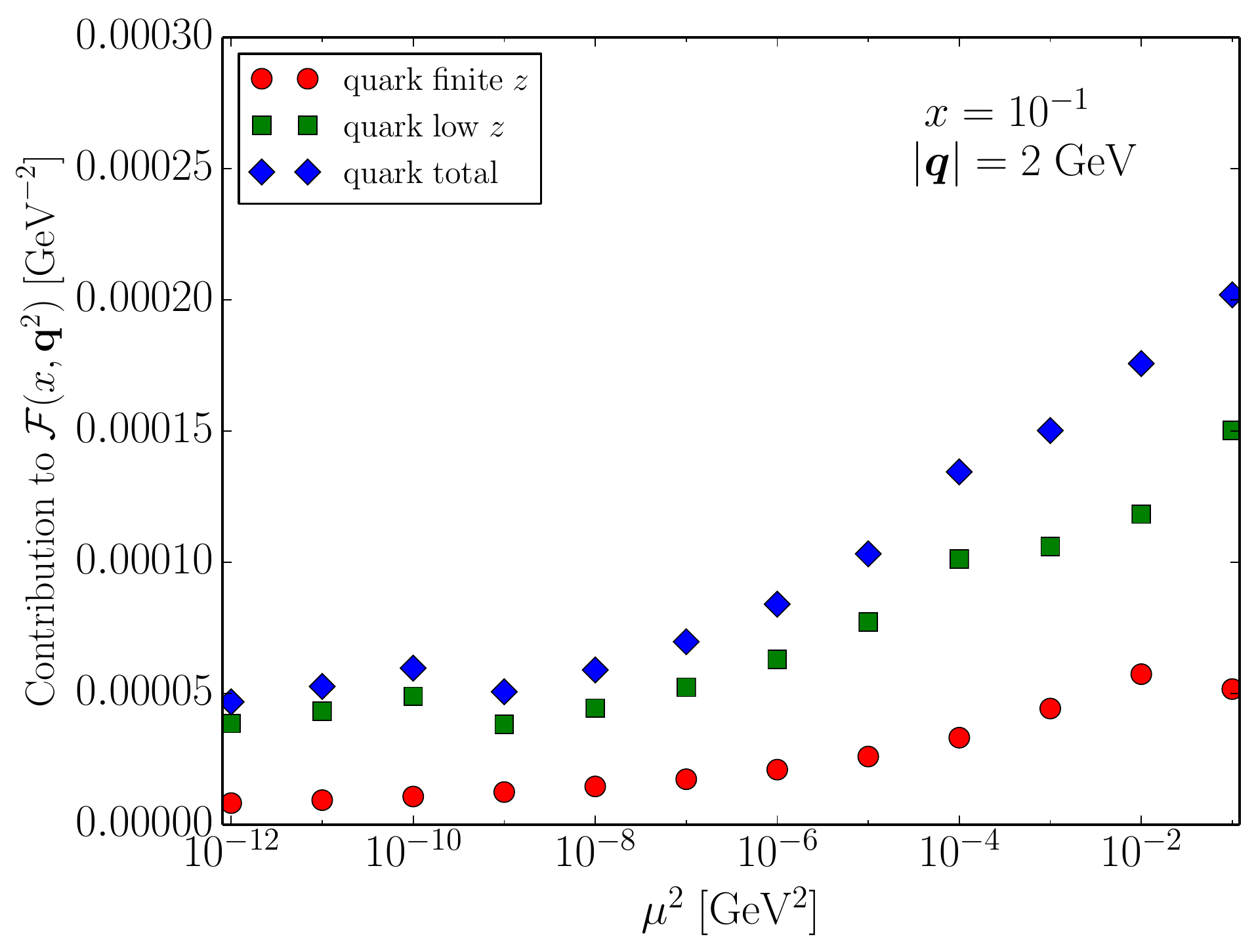}
  \hspace{15pt}
  \includegraphics[width=0.46\textwidth]{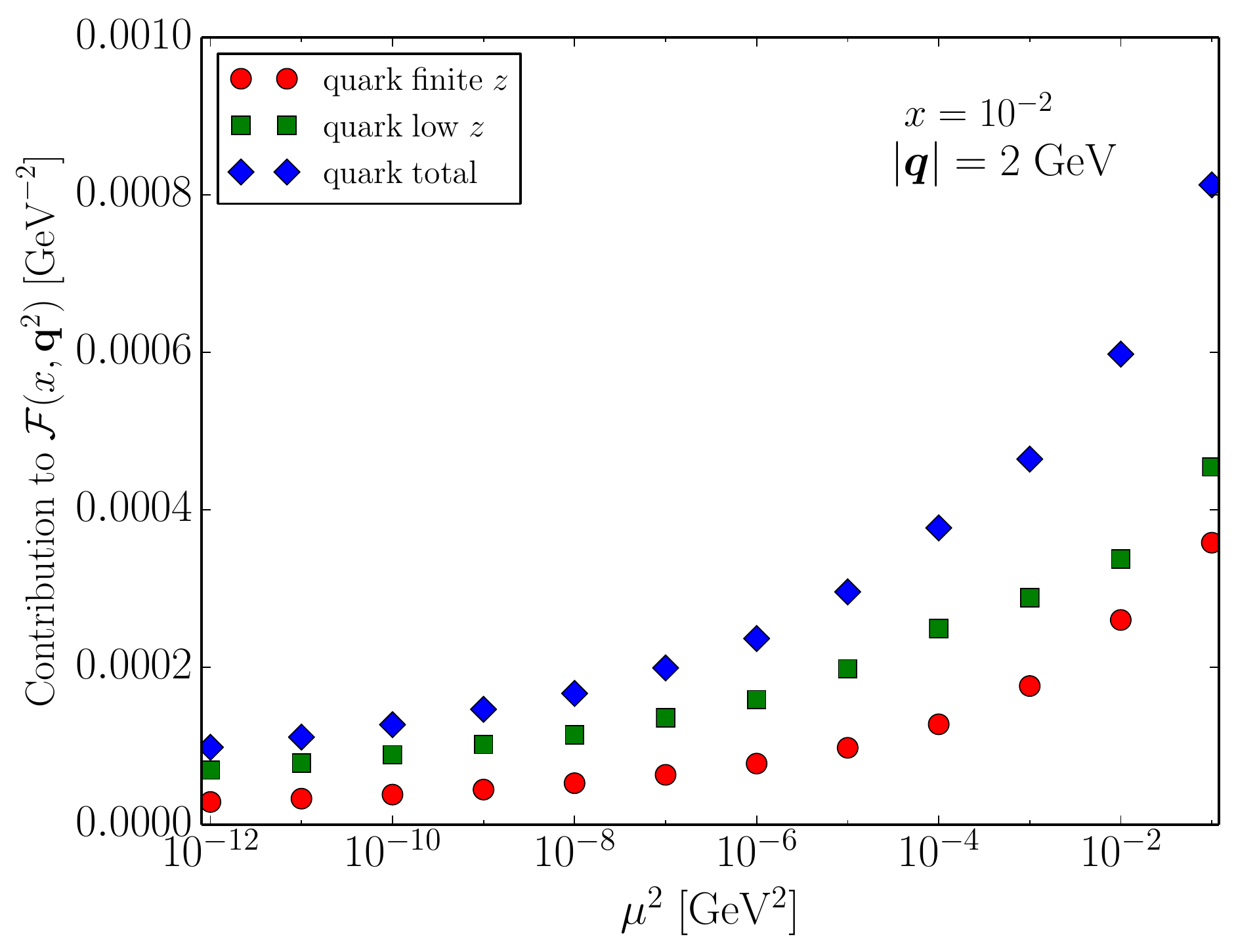}
  \caption{
  \it{The figure visualizes the cutoff dependence of the low $z$ and
  finite $z$ quark terms contributing to the gluon density.}}
  \label{fig:figurequark}
\end{figure*}

\begin{figure*}[th]
\centering
  \includegraphics[width=0.46\textwidth]{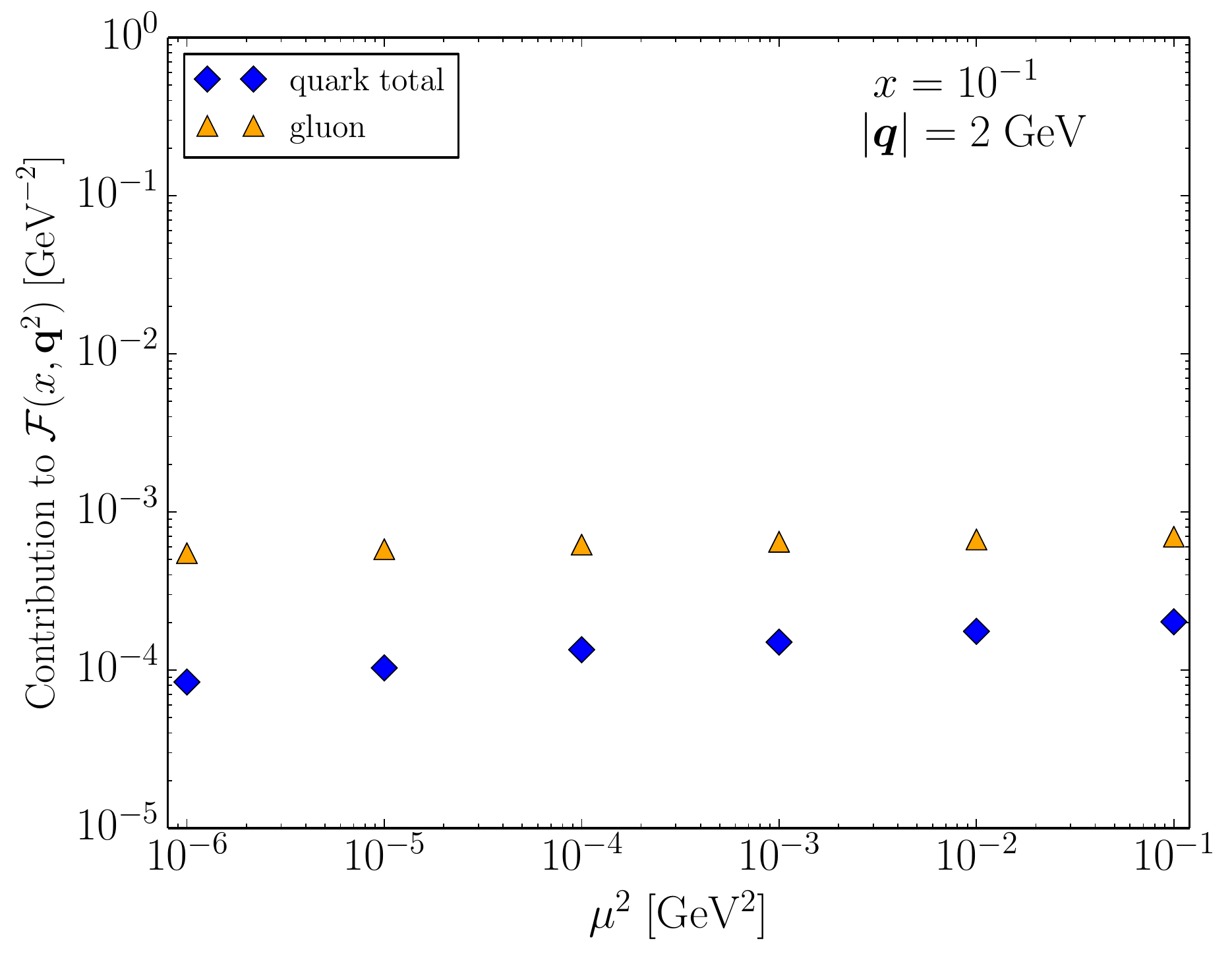}
  \hspace{15pt}
  \includegraphics[width=0.46\textwidth]{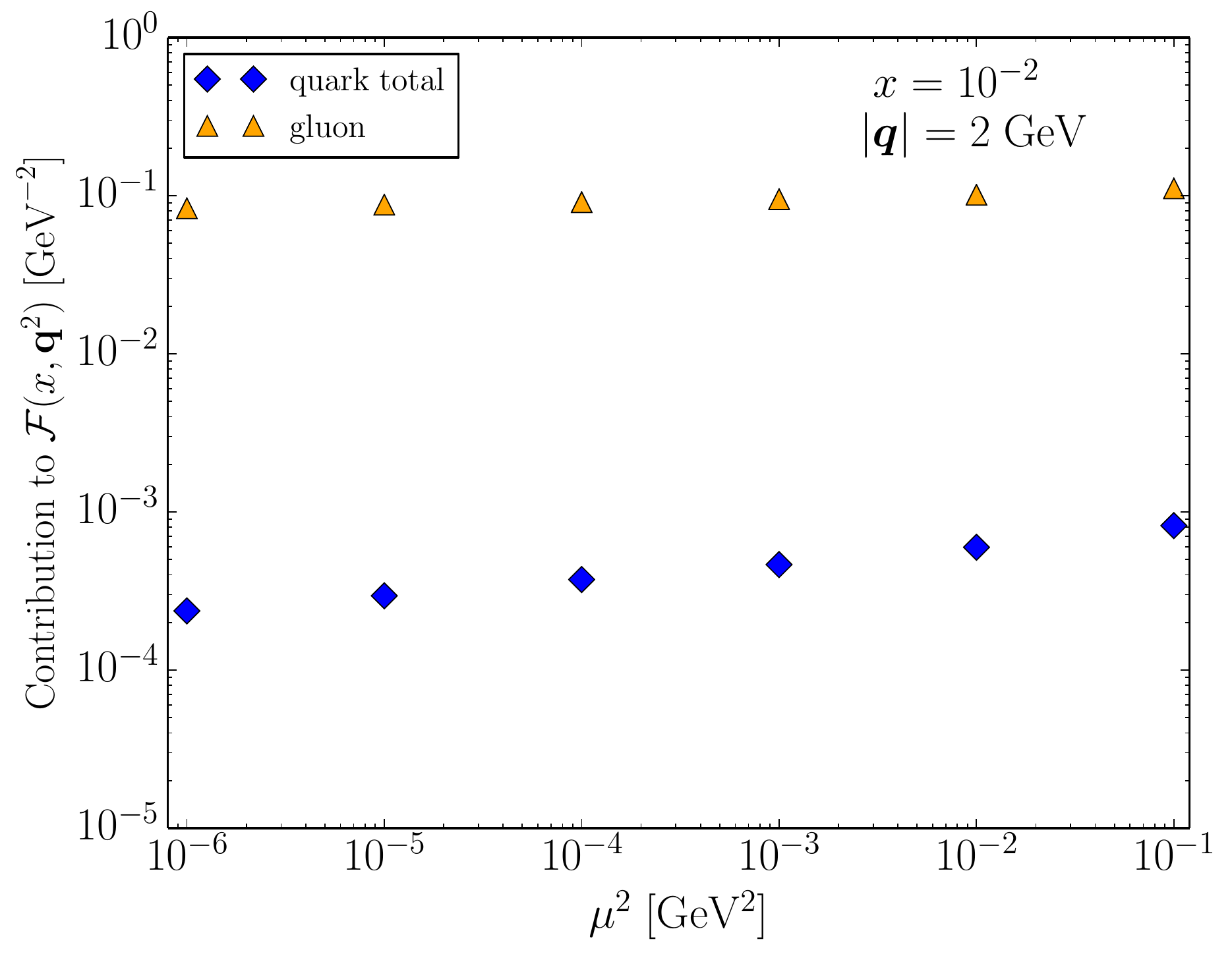}
  \caption{
  \it{The figure shows the relative contribution of gluon and quark terms at
  $|\qt|=2$ GeV to the density of gluons for $x=10^{-1}$ (left) and $x=10^{-2}$ (right)
  as a function of cutoff $\mu$.}}
  \label{fig:figurequarkgluon}
\end{figure*}

%of convergence for $\mu^2\to0$}
%===============================
First of all we would like to estimate the stability of
eq.~\eqref{eq:linear} with respect to the cutoff $\mu$. To address
this issue we need to assume some form of quark distribution
${\cal Q}$.%
\footnote{When a complete set of equations, including both quarks and
  gluons, will be available the quark distribution will be computed
  while solving this system.}  To have a realistic form of this
function we use the DLC 2016 set of parton
densities~\cite{Kutak:2016mik}.%
\footnote{The DLC 2016 PDFs are defined for $\mu>1.3$ GeV. Since in
  our case $\mu$ is a technical cut-off which extends below this
  value, we assume for this study that the quark distribution for
  $\mu<1.3$ GeV is constant.}  Employing this set of PDFs (with just
one quark flavor) we perform the convolution of the low $z$ and finite
$z$ parts of the $P_{gq}$ kernel.  We evaluate the quark density for
$x=10^{-1}$ and $10^{-2}$ and we are lead to the conclusion that as
$\mu^2\to0$ the cutoff dependence gets weaker see
Fig.~\ref{fig:figurequark}.  The other issue we can already address is
the importance of the quark contribution to gluons at the unintegrated
level as predicted by our equation.  To answer this question we
perform one iteration of the appropriate splitting functions on gluon
and quark distribution in eq.~\eqref{eq:linear}, where for simplicity
we again restrict to the case of one quark flavor. Since the
applicability of the used gluon splitting function is limited to the
low $x$ domain, we study the effects only for
% moderate ($x=10^{-1}$) and low values ($x=10^{-2}$)
moderate $x$ values $x=\{10^{-1}, 10^{-2}\}$
of the longitudinal momentum
fraction. From Fig.~\ref{fig:figurequarkgluon} we can see that at
$x=10^{-2}$ the quark contribution is much smaller than the gluon
contribution and can be neglected, while at larger $x$ it starts to be
relevant.

%===============================
\subsection{Low $x$ limit}
\label{subsec:lowX}
%===============================
Since we are already working in the low-$x$ approximation for the
gluon part it is natural to also study the result of
Sec.~\ref{subsec:comb} when a similar limit is taken in case of $P_{gq}$
kernel. To do it in a correct way we need to go back to the kernel of
eq.~\eqref{eq:Pgq} and take the $z\to0$ limit, which gives
\begin{equation}
\label{eq:Pgq_lowz}
P_{gq}(z,\pt,{\bf q})= C_F\frac{2}{z},
\end{equation}
and repeat the steps of Sec.~\ref{sec:comb}.
If we do this and substitute the simplified kernel to eq.~\eqref{eq:Mellin_q}
it is easy to check that we will obtain
%----------------
\begin{widetext}
\begin{align}
&\int\limits_{c-i\infty}^{c+i\infty}  \frac{d\omega}{2\pi i}\,
x^{-\omega}\frac{\alpha_s}{2\pi}\frac{\omega}{\omega+\overline\omega}
\int\frac{d^2{\pt}}{\pi\pt^2}\,
\int_0^1 dz z^{\omega} P_{gq}(z,\pt,\qt)
\int_0^1\,dyy^{\omega-1}{\cal Q}(y,|\qt+\pt|^2)\theta(\pt^2-\mu^2)
\nonumber\\& =
\int\limits_{c-i\infty}^{c+i\infty} \frac{d\omega}{2\pi i} \,
x^{-\omega}\frac{\alpha_s}{\pi}C_F \frac{\omega}{\omega+\overline\omega}
\int\frac{d^2{\pt}}{\pi\pt^2}\,
\int_0^1 dz z^{\omega-1}
\int_0^1\,dyy^{\omega-1}{\cal Q}(y,|\qt+\pt|^2)\theta(\pt^2-\mu^2)
\nonumber\\& =
\frac{\alpha_s}{\pi}C_F 
\int\frac{d^2{\pt}}{\pi\pt^2}\,
\int_x^1 \frac{dz}{z} z^{\overline\omega}
{\cal Q}\left(\frac{x}{z},|\qt+\pt|^2\right)\theta(\pt^2-\mu^2).
\end{align}
This leads to the final result (taking into account the nonlinear effects) in the low-$x$ limit
\begin{align}
& {\cal F}(x,\qt^2)
=
\tilde{\cal F}^0(x,\qt^2)
\nonumber\\&+
\frac{\alpha_s}{\pi} 
\int \frac{d^2 \pt}{\pi \pt^2} \, \theta(\pt^2-\mu^2)
  \int\limits_x^1 \frac{dz}{z}  \Delta_R(z, \qt^2, \mu^2) \bigg[
    C_A {\cal F}\left(\frac{x}{z},|\qt+\pt|^2\right)
  + C_F {\cal Q}\left(\frac{x}{z},|\qt+\pt|^2\right) \bigg]
% \nonumber\\&+
\nonumber\\&-
 \frac{2\alpha_s^2}{R^2}
 \int\limits_x^1\frac{dz}{z}\Delta_R(z, \qt^2, \mu^2)
\Bigg[
\left(
\int\limits_{{\qt}^2}^{\infty} \frac{\dd l^2}{l^2}
{\cal F}\left(\frac x z, l^2\right)
\right)^2+
{\cal F}\left(\frac{x}{z}, \qt^2\right)\int\limits_{\qt^2}^{\infty} \frac{\dd l^2}{l^2}
\ln \left(\frac {l^2}{{\qt}^2}\right)
{\cal F}\left(\frac{x}{z}, l^2\right)\Bigg] .
\label{eq:2}
\end{align}
\end{widetext}
%----------------
We find that when the low-$x$ limit is taken, the treatment of the
$\frac{1}{\pt^2}$ singularity of the quark-to-gluon splitting appears
to be in 1-1 correspondence with the low $x$ gluon-to-gluon splitting,
 {\it i.e.} the contributions of gluon ($\mathcal{F}$) and
  ($\mathcal{Q}$) in the second line of eq.~\eqref{eq:2} differ only
  by their overall color factor and are both regulated by a) the
  cut-off $\mu^2$ and b) the Regge form factor $\Delta_R$ which
  ensures stability in the limit $\mu \to 0$.  Note that, since the
  treatment of the quark part of the evolution equation is not
  affected by the presence of the non-linear terms in the second line
  of eq.~\eqref{eq:2}, we expect this equation to behave in the limit
  $\mu \to 0$ identical to its linear truncation. This is in
  particular true for the quark part which initially provided the main
  source of a potential instability; the purely gluonic non-linear
  evolution equation is on the other hand known to be stable in the
  limit $\mu \to 0$~\cite{Kutak:2013kfa}.

%%%%%%%%%%%%%%%%%%%%%%%%%%%%%%%%%%%%%%%
\section{Summary and outlook}
\label{sec:conclusions}
%%%%%%%%%%%%%%%%%%%%%%%%%%%%%%%%%%%%%%%
In the paper we have constructed a nonlinear TMD evolution equation
for gluons, receiving contribution from quarks. In order to regularize
the divergences of the $P_{gq}$ kernel, we introduced a cut-off and
identified the latter with a  similar cut-off introduced in the
context of  the BFKL
kernel; within the BFKL kernel this cut-off  serves 
for splitting off the low $p_T$ region of the real contributions and
adding the latter to the virtual corrections. Resumming the combined
contribution of `virtual part and low $p_T$ real part' of the BFKL
kernel to all orders
in the strong coupling,  one finds that both the pure gluonic
contribution to the evolution equation  as well as
the quark induced term are finite if we send this  cut-off to zero.
In particular we demonstrated, via performing one iteration
of the kernels, that the equation has a realistic chance to be stable against
variation of the cutoff parameter,  since after one iteration the result
stabilizes.  To perform a fully consistent study of the complete system of
$k_T$-dependent evolution equations,  we need to calculate the virtual
contributions to the quark-to-quark splitting functions in the framework of
$k_T$-factorization. We plan to address this question through a
calculation of corresponding virtual corrections, using the
calculational framework developed in
\cite{Hentschinski:2011tz,Hentschinski:2011xg,Chachamis:2012gh,Chachamis:2012mw,Chachamis:2012cc,Chachamis:2013hma}
to determine  loop-corrections within high-energy
factorization. These results will then be combined with the already
known real contributions, using an appropriate extension of the Curci
Furmanski Petronzio method to off-shell initial quarks.
We also expect that one can again perform a similar resummation of the
combined virtual and low $p_T$ real terms, leading to a system of
equations for quarks and gluons on equal footing.  Furthermore, a
numerical study of the complete system will then allow for a more
consistent study of the cutoff dependence and -- in addition -- enable
us to study the impact of nonlinearities on the quark density.

%%%%%%%%%%%%%%%%%%%%%%%%%%%%%%%%%%%%%%%
\section*{Acknowledgments}
%%%%%%%%%%%%%%%%%%%%%%%%%%%%%%%%%%%%%%%
We acknowledge useful discussions with Hannes Jung and Francesco
Hautmann at the early stage of the project.  K.K has been supported by
Narodowe Centrum Nauki with Sonata Bis grant
DEC-2013/10/E/ST2/00656. M.H. acknowledges support by CONACyT-Mexico
grant numbers CB-2014-22117 and Proy.~No. 241408.

%%%%%%%%%%%%%%%%%%%%%%%%%%%%%%%%%%%%%%%%%%%%%%%%%%
%%%%%%%%%%%%%%%%%%%%%%%%%%%%%%%%%%%%%%%%%%%%%%%%%%
\bibliographystyle{utphys}
\bibliography{refs} 

\providecommand{\href}[2]{#2}\begingroup\raggedright\begin{thebibliography}{10}

\bibitem{Angeles-Martinez:2015sea}
R.~Angeles-Martinez {\em et al.}, ``{Transverse Momentum Dependent (TMD) parton
  distribution functions: status and prospects},'' {\em Acta Phys. Polon.} {\bf
  B46} (2015), no.~12, 2501--2534,
\href{http://www.arXiv.org/abs/1507.05267}{{\tt 1507.05267}}.
%%CITATION = ARXIV:1507.05267;%%.

\bibitem{Kuraev:1976ge}
E.~A. Kuraev, L.~N. Lipatov, and V.~S. Fadin, ``{Multi - Reggeon Processes in
  the Yang-Mills Theory},'' {\em Sov. Phys. JETP} {\bf 44} (1976) 443--450.
[Zh. Eksp. Teor. Fiz.71,840(1976)].
%%CITATION = SPHJA,44,443;%%.

\bibitem{Balitsky:1978ic}
I.~I. Balitsky and L.~N. Lipatov, ``{The Pomeranchuk Singularity in Quantum
  Chromodynamics},'' {\em Sov. J. Nucl. Phys.} {\bf 28} (1978) 822--829.
[Yad. Fiz.28,1597(1978)].
%%CITATION = SJNCA,28,822;%%.

\bibitem{Catani:1990eg}
S.~Catani, M.~Ciafaloni, and F.~Hautmann, ``{High-energy factorization and
  small x heavy flavor production},'' {\em Nucl. Phys.} {\bf B366} (1991)
135--188.
%%CITATION = NUPHA,B366,135;%%.

\bibitem{Catani:1990xk}
S.~Catani, M.~Ciafaloni, and F.~Hautmann, ``{GLUON CONTRIBUTIONS TO SMALL x
  HEAVY FLAVOR PRODUCTION},'' {\em Phys. Lett.} {\bf B242} (1990)
97--102.
%%CITATION = PHLTA,B242,97;%%.

\bibitem{Dominguez:2011wm}
F.~Dominguez, C.~Marquet, B.-W. Xiao, and F.~Yuan, ``{Universality of
  Unintegrated Gluon Distributions at small x},'' {\em Phys. Rev.} {\bf D83}
  (2011) 105005,
\href{http://www.arXiv.org/abs/1101.0715}{{\tt 1101.0715}}.
%%CITATION = ARXIV:1101.0715;%%.

\bibitem{Kotko:2015ura}
P.~Kotko, K.~Kutak, C.~Marquet, E.~Petreska, S.~Sapeta, and A.~van Hameren,
  ``{Improved TMD factorization for forward dijet production in dilute-dense
  hadronic collisions},'' {\em JHEP} {\bf 09} (2015) 106,
\href{http://www.arXiv.org/abs/1503.03421}{{\tt 1503.03421}}.
%%CITATION = ARXIV:1503.03421;%%.

\bibitem{Deak:2009xt}
M.~Deak, F.~Hautmann, H.~Jung, and K.~Kutak, ``{Forward Jet Production at the
  Large Hadron Collider},'' {\em JHEP} {\bf 09} (2009) 121,
\href{http://www.arXiv.org/abs/0908.0538}{{\tt 0908.0538}}.
%%CITATION = ARXIV:0908.0538;%%.

\bibitem{vanHameren:2012if}
A.~van Hameren, P.~Kotko, and K.~Kutak, ``{Helicity amplitudes for high-energy
  scattering},'' {\em JHEP} {\bf 01} (2013) 078,
\href{http://www.arXiv.org/abs/1211.0961}{{\tt 1211.0961}}.
%%CITATION = ARXIV:1211.0961;%%.

\bibitem{vanHameren:2012uj}
A.~van Hameren, P.~Kotko, and K.~Kutak, ``{Multi-gluon helicity amplitudes with
  one off-shell leg within high energy factorization},'' {\em JHEP} {\bf 12}
  (2012) 029,
\href{http://www.arXiv.org/abs/1207.3332}{{\tt 1207.3332}}.
%%CITATION = ARXIV:1207.3332;%%.

\bibitem{Hautmann:2012rf}
F.~Hautmann, M.~Hentschinski, and H.~Jung, ``{Unintegrated sea quark at small x
  and vector boson production},'' in {\em {Proceedings, 47th Rencontres de
  Moriond on QCD and High Energy Interactions}}, pp.~391--394.
\newblock 2012.
\newblock
\href{http://www.arXiv.org/abs/1209.6305}{{\tt 1209.6305}}.
\newblock
%%CITATION = ARXIV:1209.6305;%%.

\bibitem{Hautmann:2014uua}
F.~Hautmann, H.~Jung, and S.~T. Monfared, ``{The CCFM uPDF evolution uPDFevolv
  Version 1.0.00},'' {\em Eur. Phys. J.} {\bf C74} (2014) 3082,
\href{http://www.arXiv.org/abs/1407.5935}{{\tt 1407.5935}}.
%%CITATION = ARXIV:1407.5935;%%.

\bibitem{Gituliar:2015agu}
O.~Gituliar, M.~Hentschinski, and K.~Kutak, ``{Transverse-momentum-dependent
  quark splitting functions in $k_{T}-$factorization: real contributions},''
  {\em JHEP} {\bf 01} (2016) 181,
\href{http://www.arXiv.org/abs/1511.08439}{{\tt 1511.08439}}.
%%CITATION = ARXIV:1511.08439;%%.

\bibitem{Catani:1994sq}
S.~Catani and F.~Hautmann, ``{High-energy factorization and small x deep
  inelastic scattering beyond leading order},'' {\em Nucl. Phys.} {\bf B427}
  (1994) 475--524,
\href{http://www.arXiv.org/abs/hep-ph/9405388}{{\tt hep-ph/9405388}}.
%%CITATION = HEP-PH/9405388;%%.

\bibitem{Hautmann:2012sh}
F.~Hautmann, M.~Hentschinski, and H.~Jung, ``{Forward Z-boson production and
  the unintegrated sea quark density},'' {\em Nucl. Phys.} {\bf B865} (2012)
  54--66,
\href{http://www.arXiv.org/abs/1205.1759}{{\tt 1205.1759}}.
%%CITATION = ARXIV:1205.1759;%%.

\bibitem{Curci:1980uw}
G.~Curci, W.~Furmanski, and R.~Petronzio, ``{Evolution of Parton Densities
  Beyond Leading Order: The Nonsinglet Case},'' {\em Nucl. Phys.} {\bf B175}
  (1980)
27--92.
%%CITATION = NUPHA,B175,27;%%.

\bibitem{Jadach:2011kc}
S.~Jadach, A.~Kusina, M.~Skrzypek, and M.~Slawinska, ``{Two real parton
  contributions to non-singlet kernels for exclusive QCD DGLAP evolution},''
  {\em JHEP} {\bf 08} (2011) 012,
\href{http://www.arXiv.org/abs/1102.5083}{{\tt 1102.5083}}.
%%CITATION = ARXIV:1102.5083;%%.

\bibitem{Gituliar:2014eba}
O.~Gituliar, S.~Jadach, A.~Kusina, and M.~Skrzypek, ``{On regularizing the
  infrared singularities in QCD NLO splitting functions with the new Principal
  Value prescription},'' {\em Phys. Lett.} {\bf B732} (2014) 218--222,
\href{http://www.arXiv.org/abs/1401.5087}{{\tt 1401.5087}}.
%%CITATION = ARXIV:1401.5087;%%.

\bibitem{Jadach:2016zgk}
S.~Jadach, A.~Kusina, W.~Placzek, and M.~Skrzypek, ``{On the dependence of QCD
  splitting functions on the choice of the evolution variable},'' {\em JHEP}
  {\bf 08} (2016) 092,
\href{http://www.arXiv.org/abs/1606.01238}{{\tt 1606.01238}}.
%%CITATION = ARXIV:1606.01238;%%.

\bibitem{Ciafaloni:1987ur}
M.~Ciafaloni, ``{Coherence Effects in Initial Jets at Small q**2 / s},'' {\em
  Nucl. Phys.} {\bf B296} (1988)
49--74.
%%CITATION = NUPHA,B296,49;%%.

\bibitem{Catani:1989sg}
S.~Catani, F.~Fiorani, and G.~Marchesini, ``{Small x Behavior of Initial State
  Radiation in Perturbative QCD},'' {\em Nucl. Phys.} {\bf B336} (1990)
18--85.
%%CITATION = NUPHA,B336,18;%%.

\bibitem{Catani:1989yc}
S.~Catani, F.~Fiorani, and G.~Marchesini, ``{QCD Coherence in Initial State
  Radiation},'' {\em Phys. Lett.} {\bf B234} (1990)
339--345.
%%CITATION = PHLTA,B234,339;%%.

\bibitem{Marchesini:1994wr}
G.~Marchesini, ``{QCD coherence in the structure function and associated
  distributions at small x},'' {\em Nucl. Phys.} {\bf B445} (1995) 49--80,
\href{http://www.arXiv.org/abs/hep-ph/9412327}{{\tt hep-ph/9412327}}.
%%CITATION = HEP-PH/9412327;%%.

\bibitem{Kimber:1999xc}
M.~A. Kimber, A.~D. Martin, and M.~G. Ryskin, ``{Unintegrated parton
  distributions and prompt photon hadroproduction},'' {\em Eur. Phys. J.} {\bf
  C12} (2000) 655--661,
\href{http://www.arXiv.org/abs/hep-ph/9911379}{{\tt hep-ph/9911379}}.
%%CITATION = HEP-PH/9911379;%%.

\bibitem{Kimber:2001sc}
M.~A. Kimber, A.~D. Martin, and M.~G. Ryskin, ``{Unintegrated parton
  distributions},'' {\em Phys. Rev.} {\bf D63} (2001) 114027,
\href{http://www.arXiv.org/abs/hep-ph/0101348}{{\tt hep-ph/0101348}}.
%%CITATION = HEP-PH/0101348;%%.

\bibitem{Schafer:2016qmk}
W.~Schafer and A.~Szczurek, ``{Low mass Drell-Yan production of lepton pairs at
  forward directions at the LHC: a hybrid approach},'' {\em Phys. Rev.} {\bf
  D93} (2016), no.~7, 074014,
\href{http://www.arXiv.org/abs/1602.06740}{{\tt 1602.06740}}.
%%CITATION = ARXIV:1602.06740;%%.

\bibitem{Maciula:2015kea}
R.~Maciuła, A.~Szczurek, and M.~Łuszczak, ``{Open charm meson production at
  BNL RHIC within $k_{t}$-factorization approach and revision of their
  semileptonic decays},'' {\em Phys. Rev.} {\bf D92} (2015), no.~5, 054006,
\href{http://www.arXiv.org/abs/1505.05038}{{\tt 1505.05038}}.
%%CITATION = ARXIV:1505.05038;%%.

\bibitem{Ducloue:2015jba}
B.~Ducloué, L.~Szymanowski, and S.~Wallon, ``{Evaluating the double parton
  scattering contribution to Mueller-Navelet jets production at the LHC},''
  {\em Phys. Rev.} {\bf D92} (2015), no.~7, 076002,
\href{http://www.arXiv.org/abs/1507.04735}{{\tt 1507.04735}}.
%%CITATION = ARXIV:1507.04735;%%.

\bibitem{Nefedov:2013ywa}
M.~A. Nefedov, V.~A. Saleev, and A.~V. Shipilova, ``{Dijet azimuthal
  decorrelations at the LHC in the parton Reggeization approach},'' {\em Phys.
  Rev.} {\bf D87} (2013), no.~9, 094030,
\href{http://www.arXiv.org/abs/1304.3549}{{\tt 1304.3549}}.
%%CITATION = ARXIV:1304.3549;%%.

\bibitem{Baranov:2016mix}
S.~P. Baranov, A.~V. Lipatov, M.~A. Malyshev, A.~M. Snigirev, and N.~P. Zotov,
  ``{Associated production of electroweak bosons and heavy mesons at LHCb and
  the prospects to observe double parton interactions},'' {\em Phys. Rev.} {\bf
  D93} (2016), no.~9, 094013,
\href{http://www.arXiv.org/abs/1604.03025}{{\tt 1604.03025}}.
%%CITATION = ARXIV:1604.03025;%%.

\bibitem{Kwiecinski:1996td}
J.~Kwiecinski, A.~D. Martin, and P.~J. Sutton, ``{Constraints on gluon
  evolution at small x},'' {\em Z. Phys.} {\bf C71} (1996) 585--594,
\href{http://www.arXiv.org/abs/hep-ph/9602320}{{\tt hep-ph/9602320}}.
%%CITATION = HEP-PH/9602320;%%.

\bibitem{Kutak:2011fu}
K.~Kutak, K.~Golec-Biernat, S.~Jadach, and M.~Skrzypek, ``{Nonlinear equation
  for coherent gluon emission},'' {\em JHEP} {\bf 02} (2012) 117,
\href{http://www.arXiv.org/abs/1111.6928}{{\tt 1111.6928}}.
%%CITATION = ARXIV:1111.6928;%%.

\bibitem{Kutak:2012qk}
K.~Kutak, ``{Resummation in nonlinear equation for high energy factorisable
  gluon density and its extension to include coherence},'' {\em JHEP} {\bf 12}
  (2012) 033,
\href{http://www.arXiv.org/abs/1206.5757}{{\tt 1206.5757}}.
%%CITATION = ARXIV:1206.5757;%%.

\bibitem{Gribov:1984tu}
L.~V. Gribov, E.~M. Levin, and M.~G. Ryskin, ``{Semihard Processes in QCD},''
  {\em Phys. Rept.} {\bf 100} (1983)
1--150.
%%CITATION = PRPLC,100,1;%%.

\bibitem{Gelis:2010nm}
F.~Gelis, E.~Iancu, J.~Jalilian-Marian, and R.~Venugopalan, ``{The Color Glass
  Condensate},'' {\em Ann. Rev. Nucl. Part. Sci.} {\bf 60} (2010) 463--489,
\href{http://www.arXiv.org/abs/1002.0333}{{\tt 1002.0333}}.
%%CITATION = ARXIV:1002.0333;%%.

\bibitem{Kutak:2003bd}
K.~Kutak and J.~Kwiecinski, ``{Screening effects in the ultrahigh-energy
  neutrino interactions},'' {\em Eur. Phys. J.} {\bf C29} (2003) 521,
\href{http://www.arXiv.org/abs/hep-ph/0303209}{{\tt hep-ph/0303209}}.
%%CITATION = HEP-PH/0303209;%%.

\bibitem{Bartels:2007dm}
J.~Bartels and K.~Kutak, ``{A Momentum Space Analysis of the Triple Pomeron
  Vertex in pQCD},'' {\em Eur. Phys. J.} {\bf C53} (2008) 533--548,
\href{http://www.arXiv.org/abs/0710.3060}{{\tt 0710.3060}}.
%%CITATION = ARXIV:0710.3060;%%.

\bibitem{Kovchegov:1999yj}
Y.~V. Kovchegov, ``{Small x F(2) structure function of a nucleus including
  multiple pomeron exchanges},'' {\em Phys. Rev.} {\bf D60} (1999) 034008,
\href{http://www.arXiv.org/abs/hep-ph/9901281}{{\tt hep-ph/9901281}}.
%%CITATION = HEP-PH/9901281;%%.

\bibitem{Balitsky:1995ub}
I.~Balitsky, ``{Operator expansion for high-energy scattering},'' {\em Nucl.
  Phys.} {\bf B463} (1996) 99--160,
\href{http://www.arXiv.org/abs/hep-ph/9509348}{{\tt hep-ph/9509348}}.
%%CITATION = HEP-PH/9509348;%%.

\bibitem{Kutak:2016mik}
K.~Kutak, R.~Maciula, M.~Serino, A.~Szczurek, and A.~van Hameren, ``{Four-jet
  production in single- and double-parton scattering within high-energy
  factorization},'' {\em JHEP} {\bf 04} (2016) 175,
\href{http://www.arXiv.org/abs/1602.06814}{{\tt 1602.06814}}.
%%CITATION = ARXIV:1602.06814;%%.

\bibitem{Kutak:2013kfa}
K.~Kutak, W.~Płaczek, and D.~Toton, ``{Numerical solution of the integral form
  of the resummed Balitsky-Kovchegov equation},'' {\em Acta Phys. Polon.} {\bf
  B44} (2013), no.~7, 1527--1535,
\href{http://www.arXiv.org/abs/1303.0431}{{\tt 1303.0431}}.
%%CITATION = ARXIV:1303.0431;%%.

\bibitem{Hentschinski:2011tz}
M.~Hentschinski and A.~Sabio~Vera, ``{NLO jet vertex from Lipatov's QCD
  effective action},'' {\em Phys. Rev.} {\bf D85} (2012) 056006,
\href{http://www.arXiv.org/abs/1110.6741}{{\tt 1110.6741}}.
%%CITATION = ARXIV:1110.6741;%%.

\bibitem{Hentschinski:2011xg}
M.~Hentschinski, ``{Pole prescription of higher order induced vertices in
  Lipatov's QCD effective action},'' {\em Nucl. Phys.} {\bf B859} (2012)
  129--142,
\href{http://www.arXiv.org/abs/1112.4509}{{\tt 1112.4509}}.
%%CITATION = ARXIV:1112.4509;%%.

\bibitem{Chachamis:2012gh}
G.~Chachamis, M.~Hentschinski, J.~D. Madrigal~Martinez, and A.~Sabio~Vera,
  ``{Quark contribution to the gluon Regge trajectory at NLO from the high
  energy effective action},'' {\em Nucl. Phys.} {\bf B861} (2012) 133--144,
\href{http://www.arXiv.org/abs/1202.0649}{{\tt 1202.0649}}.
%%CITATION = ARXIV:1202.0649;%%.

\bibitem{Chachamis:2012mw}
G.~Chachamis, M.~Hentschinski, J.~D. Madrigal~Martinez, and A.~Sabio~Vera,
  ``{Forward jet production \& quantum corrections to the gluon Regge
  trajectory from Lipatov`s high energy effective action},'' {\em Phys. Part.
  Nucl.} {\bf 45} (2014), no.~4, 788--799,
\href{http://www.arXiv.org/abs/1211.2050}{{\tt 1211.2050}}.
%%CITATION = ARXIV:1211.2050;%%.

\bibitem{Chachamis:2012cc}
G.~Chachamis, M.~Hentschinski, J.~D. Madrigal~Martinez, and A.~Sabio~Vera,
  ``{Next-to-leading order corrections to the gluon-induced forward jet vertex
  from the high energy effective action},'' {\em Phys. Rev.} {\bf D87} (2013),
  no.~7, 076009,
\href{http://www.arXiv.org/abs/1212.4992}{{\tt 1212.4992}}.
%%CITATION = ARXIV:1212.4992;%%.

\bibitem{Chachamis:2013hma}
G.~Chachamis, M.~Hentschinski, J.~D. Madrigal~Martinez, and A.~Sabio~Vera,
  ``{Gluon Regge trajectory at two loops from Lipatov's high energy effective
  action},'' {\em Nucl. Phys.} {\bf B876} (2013) 453--472,
\href{http://www.arXiv.org/abs/1307.2591}{{\tt 1307.2591}}.
%%CITATION = ARXIV:1307.2591;%%.

\end{thebibliography}\endgroup
%%%%%%%%%%%%%%%%%%%%%%%%%%%%%%%%%%%%%%%%%%%%%%%%%%

\end{document}